\documentclass{osa-article}

\journal{boe}

\articletype{}

\begin{document}

\title{Non-invasive optical measurement of arterial blood flow speed}

\author{Alex Ce Zhang,\authormark{1} Yu-Hwa Lo\authormark{1*} }

\address{\authormark{1}Department of Electrical and Computer Engineering, University of California San Diego,  9500 Gilman Driver, La Jolla ,CA, 92092, USA}

\email{\authormark{*}ylo@ucsd.edu} 



\begin{abstract}
Non-invasive measurement of the arterial blood speed gives rise to important health information such as cardio output and blood supplies to vital organs.  The magnitude and change in arterial blood speed are key indicators of the health conditions and development and progression of diseases.  We demonstrated a simple technique to directly measure the blood flow speed in main arteries based on the diffused light model.  The concept is demonstrated with a phantom that uses intralipid hydrogel to model the biological tissue and an embedded glass tube with flowing human blood to model the blood vessel.  The correlation function of the measured photocurrent was used to find the electrical field correlation function via the Siegert relation.  We have shown that the characteristic decorrelation rate (i.e. the inverse of the decoherent time) is linearly proportional to the blood speed and independent of the tube diameter.  This striking property can be explained by an approximate analytic solution for the diffused light equation in the regime where the convective flow is the dominating factor for decorrelation.  As a result, we have demonstrated a non-invasive method of measuring arterial blood speed without any prior knowledge or assumption about the geometric or mechanic properties of the blood vessels.
\end{abstract}

\section{Introduction}
Oxygenated red blood cells in the blood flow deliver essential nutrients and oxygen to organs and limbs to maintain their homeostatic conditions and proper functions. Oximeter detects the oxygen saturation level of red blood cells by using wavelength dependent absorption for oxygenated and deoxygenated blood \cite{wang2018monitoring}. Such measurements can be done with wearable devices such as smart watches. On the other hand, there has been no convenient and non-invasive method to measure the amount of blood supply or blood speed in major arteries.

Reduced blood supply is usually a sign of diseases and medical conditions. Dehydration, thromboses formation, and other physiological and pathological conditions can cause short term or long term changes in the blood speed at a given location of the artery \cite{stewart2019investigation}. The information helps early diagnosis and monitoring of diseases with impaired blood flow \cite{jayanthy2011measuring,poelma2012accurate,blumgart1927studies}. Therefore, it is desirable to directly measure the travelling velocity of red blood cells at well-defined locations such as the carotidal artery to the head, femoral artery to the lower limb, brachial artery to the upper arm, spinal arteries to the spinal cord, renal arteries to the kidneys, hepatic artery to the liver, and pulmonary artery to the lungs, etc. Since the arterial blood speed at the well-defined positions provide direct and unambiguous information about blood supply to the sites of health concerns.  

Doppler frequency shift of ultrasonic waves offers a non-invasive measurement of the blood flow speed at the well-defined position. However, acoustic measurements require close physical contacts of the ultrasound transducers with the tissue to allow efficient coupling of the acoustic wave, and gel is needed to assure the quality and reliability of the contact. Furthermore, the measured signal is highly dependent on the angle between the probe and the blood flow direction. The most convenient and ergonomic direction is to have the transducer perpendicular to the blood vessel. However, this arrangement will produce no Doppler shift signal, thus making the measurement less convenient in a homecare or nursing home setting for self-administered operation\cite{pietrangelo2018wearable}. In this regard, an optical method is potentially more desirable because a light wave can enter the tissue via free space coupling and once entering the tissue, as we will discuss later, it travels diffusively. In addition, the optical hardware including semiconductor light sources and detectors are more compact than acoustic transmitters and receivers. 
People have used dynamic scattering of laser light by red blood cells to measure the speed for blood vessels near the tissue surface \cite{kashima1992model,kashima1993study}.  Similar to the Doppler effect of sound propagation, the frequency of the scattered light by moving particles is also shifted. Although the operation principle seems to be straightforward, the physical model of the ``optical Doppler device'' is based on a key assumption that the light is only scattered once by a moving object (e.g. a traveling red blood cell) and the rest of the scatterings is produced by quasi static objects such as skin and fat.  This assumption may be reasonable for probing near surface tissue blood flow, but not for major arteries where around 40\% of the volume of the blood vessel with a diameter of a few millimeters is occupied by traveling red blood cells. Under the approximation of single light scattering by moving particles, the detected scattered light will display a frequency shift proportional to the velocity of the particle and the amount of frequency shift can be detected using the homodyne coherent detection technique. However, the above technique cannot be applied to multiple scatterings by a large number of moving particles. In such situations, the frequency spectrum of the light wave is broadened, making the interpretation of signal difficult. Because of this limit, the current technique of optical Doppler shift has been limited to measuring the blood flow in microflow channels near the skin although the blood supply by major arteries produces more valuable information for health and disease conditions.  In addition, the laser Doppler flowmetry can only measure relative speed. \cite{rajan2009review}.  
To measure the blood flow speed in major arteries, we extend the technique of optical scattering method by taking into account the effect of multiple scattering by red blood cells.  Instead of measuring the Doppler shift in the optical frequency, we measure the decorrelation time of the reflected light. The key results of our work are (a) we have shown experimentally and theoretically that one can measure the arterial blood flow velocity using a simple optical setup comprised of a single semiconductor laser and a single detector, both coupled to a multi-mode fiber bundle, and (b) we discovered that for main arteries where the diffused light model applies, the decorrelation time is inversely proportional to the RBC speed and we can measure its value without any prior knowledge about the anatomy and tissue mechanic properties of the artery.
Once entering the tissue, light has a very short scattering mean free path and the light becomes diffusive. Modelling of light propagation in this regime has led to the development of oximeter back in 70s \cite{ishimaru1978wave} and many researchers have utilized the diffused light to construct images through highly diffusive biological media, such as reconstructed breast cancer images \cite{wang2012biomedical}. The standard configuration for diffused light correlation spectroscopy includes a point source and a point detector, and the measured correlation function depends on the relative position between the source and the detector\cite{durduran2010diffuse,vishwanath2019diffuse,zanfardino2018sensitivity}. The method of diffused light correlation spectroscopy has been used in blood perfusion measurement \cite{boas2016establishing}, including brain circulations \cite{durduran2014diffuse,torricelli2014neurophotonics,verdecchia2015assessment}. The setup uses a single-mode fiber coupled detector to allow single-mode transmission of two orthogonal light polarizations\cite{rivcka1993dynamic}. The single spatial mode yields the best signal-to-noise ratio since it detects one speckle. However, this setup produces very low light intensity and requires a single photon detector (e.g. a single-photon avalanche detector (SPAD) or a photomultiplier tube (PMT)), making the setup quite sophisticated and subject to interference and stray light. 

To make the system robust and easy to operate, in this paper we use multimode fiber and a regular photodetector to detect temporal fluctuations of the light intensity. In the following we first present our experimental setup and measurement results, and then describe the physical model to show how the measured results can be used to find the red blood cell velocity.  We will show that in a simple and robust setup that is relatively insensitive to the optical alignment between the fiber bundle and the tissue, where we can reliably measure the absolute value of the traveling speed of red blood cells in a blood vessel that has a millimeter diameter embedded in a layer of tissue.  The results offer a promising solution for non-invasive measurement of the blood supply into organs and limbs.

\section{Experimental Setup}

The experimental setup and measured results are first presented here, followed by the theoretical analysis in the later sections.

\begin{figure}[h!]
\centering\includegraphics[width=13cm]{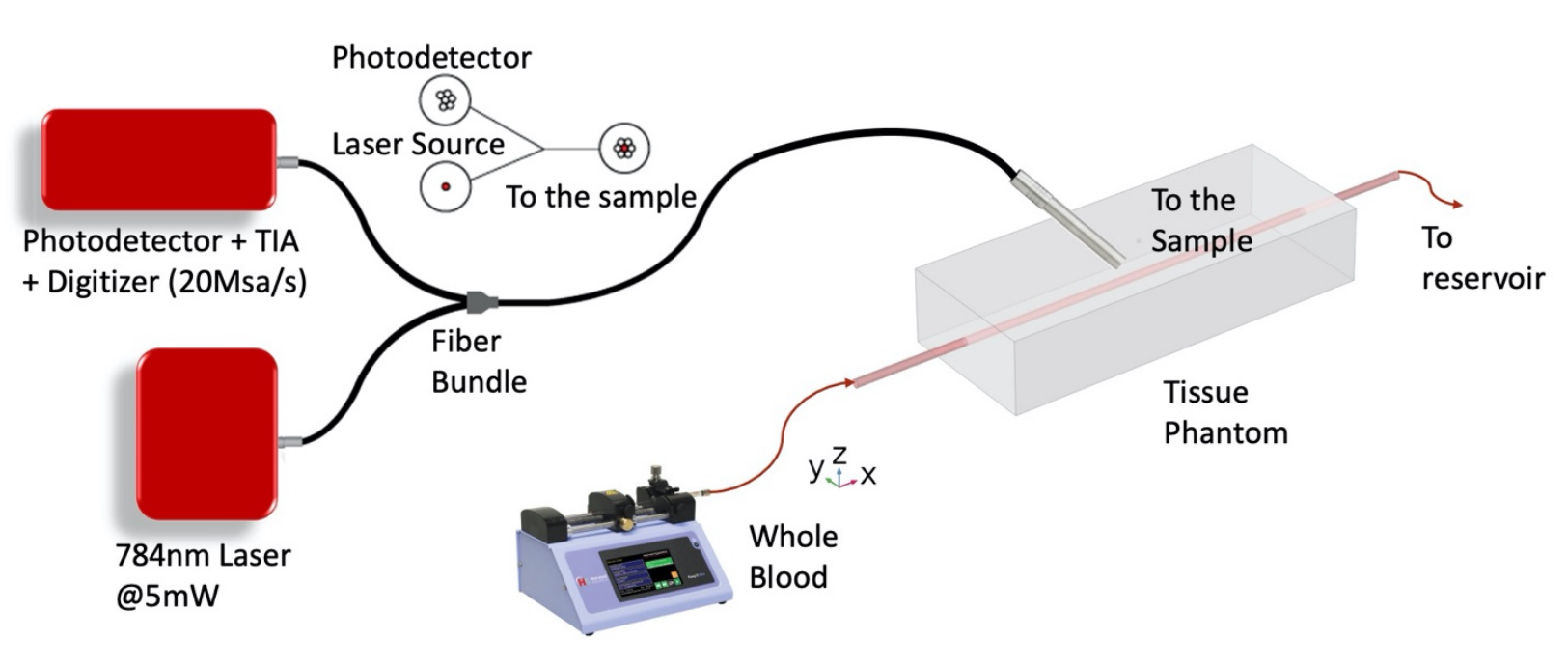}
\caption{. Experimental Setup. The laser light illuminates the tissue phantom surface at a 45-degree angle and the scattered diffused light is collected through the reflection probe fiber bundle which is coupled to a photoreceiver. The glass tube that emulates a blood vessel is embedded in the tissue phantom 5mm below the surface.}
\end{figure}
\subsection{Tissue Phantom Fabrication}
Wwe used intralipid as the scattering agent to reproduce keymimic the tissue by making its optical properties of tissue to make the scattering coefficient $\mu_s$ and anisotropy factor $g$ similar to the the values of real tissue.  Because of its similar optical properties to the bilipid membrane of cells, intralipid is commonly used to simulate tissue scattering \cite{lai2014dependence,pogue2006review,flock1992optical}. In addition, the absorption coefficient of intralipid is low and its refractive index is close to that of soft tissue\cite{ding2005determination}.  The tissue phantom is created using intralipid (Sigma-Aldrich 20\%) in gelatin gel. One percent concentration of intralipid was chosen to achieve a scattering coefficient of around 10$cm^{-1}$ at 784 $nm$. The phantom was initially prepared at 90$^{\circ}C$, and then poured into a mold where a glass tube was pre-inserted at a depth of 5mm from the phantom surface (measured from the center of the tube).The liquid phantom was immediately placed into a freezer at -18 $^{\circ}C$ for 30 minutes for rapid solidification to prevent sedimentation of intralipid to achieve a uniform scattering property. Then the sample was placed in a refrigerator at 6 $^{\circ}C$ for 30 minutes to further solidify. The subsequent experiment was usually completed within 30 minutes after the phantom fabrication to prevent evaporation induced dehydration of the phantom, which could reduce the surface height. Glass tubes of three different inner diameters were used to mimic the arteries. Tygon tubing was connected to both ends of the glass tube, with one end connected to a 10$mL$ syringe mounted on a syringe pump and the other end connected to a reservoir (See figure 1).
\subsection{Blood Preparation}
12 $mL$ of human whole blood collected on the same day of the experiment was purchased from the San Diego Blood Bank. EDTA was added to the blood sample to prevent coagulation.
\subsection{Optical and electronics setup}
We used an off-the-shelf multi-core reflection fiber bundle probe (Model:RP21 from Thorlabs) that consists of one center core surrounded by 6 peripheral cores to couple the input and reflected light. The center core was used to deliver the laser light to the blood vessel and the 6 peripheral cores that are merged into a single output are coupled to the photodetector (see Fig. 1). The fiber bundle was mounted about 45 degrees relative to the surface of the tissue phantom to prevent specular refection from the tissue surface. The fiber bundle was coarsely aligned to the position of the glass tube in the tissue phantom. The laser used in the experiment was an off-the-shelf fiber coupled laser diode at 785nm, biased slightly above its threshold current to produce an output of  5 mW. The output of the detection fiber was coupled to a silicon APD with an integrated transimpedance amplifier (Thorlabs APD410A) to achieve a transimpedance gain of 500 kV/A at 10MHz bandwidth. The APD multiplication gain was set to be the lowest ($M\sim 10$). The output signal from the APD photoreceiver entered a data acquisition board (Advantech PCI-E 1840) at a sampling rate of 20 Msps, yielding a Nyquist bandwidth of 10 MHz. The total data acquisition time for the measurement was 25 seconds. We treated each 5ms duration as one recording section, so 25 seconds of measurement produced 5,000 sections for analysis and noise cancellation. In all the data presented next, we used the average data over 25 seconds for the mean and over 1 second to determine the variations shown in error bars. 

We tested blood flows with three tube inner diameters: 0.93mm, 1.14mm, and 1.68mm. The tube diameters were measured by a digital microscope to assure high accuracy.  For each tube diameter, we flow the blood at different speeds. 
\subsection{Experimental Results}
We have used the setup in Figure 1 to measure the photocurrent correlation defined as $ \frac{\langle i(t) i(t+\tau) \rangle}{\langle i(t)\rangle^2}$where i(t) is the photocurrent at time``t'' and $\langle \rangle$ is the ensemble average.  For a 25 second measurement that can be divided into 5,000 5ms sections, the ensemble average is the average of these 5,000 sections. $\tau$ is the time delay between the instantaneous photocurrents, and is the variable for the correlation function.

One important relation is to convert the photocurrent correlation into the normalized electrical field correlation function $g_1$ represented by Eq. (1):
\begin{equation}
    g_1(\tau) = \frac{\langle E(t)E^*(t+\tau) \rangle}{\langle E(t)E^*(t)\rangle}
\end{equation}
\\Using the Siegert Relation\cite{cummins1970iii}, we can obtain a relation between the magnitude of $g_1(\tau)$ and the photocurrent correlation as shown in Eq. (2):
\begin{equation}
    \frac{\langle i(t) i(t+\tau)\rangle}{\langle i(t)\rangle^2} = 1+\frac{1}{N}|g_1(\tau)|^2+\frac{e\delta(t)}{\langle i(t)\rangle}
\end{equation}
where $i(t)$ is the measured photocurrent, N is the number of spatial modes coupled into the detection fiber bundle and collected by the detector, $\delta(t)$ is a delta function, and $e$ is the electron charge.  The last term in Eq. (2) represents the shot noise.

The magnitude of electrical field correlation $g_1$ under different flow speeds of blood is shown in semi-log plots in Figure 2-4(a), where each figure shows measurements performed with a different tube diameter. The x-axis of each plot is the logarithmic of $\tau$ in the correlation function.  From these plots we made two interesting observations: (a) At relatively low blood flow speed, the curve shows a characteristic analogous to the ``Fermi-Dirac distribution function'' if we treat $\ln (\tau)$ as ``energy''; (b) at high blood speed, the curve behaves like a superposition of two Fermi-Dirac distribution functions, one at lower ``energy'' and another at higher ``energy''.  These characteristics become more apparent if we take the derivative of $g_1$ with $\ln (\tau)$, showing that the ``Fermi level'', $\ln (T_F)$, occurs at the inflection point where the magnitude of slope reaches the maximum. 

We will discuss the physics of such characteristics in the later section.  By examining the interesting features of the measured electrical field correlation $g_1$, we can extract the ``Fermi energy, $\epsilon_F$'' or $\ln (T_F)$. In the low blood speed regime where the$|g_1|$ plot shows two superimposed Fermi-like functions, we chose the inflection point of the lower energy function (i.e. in the regime of smaller $\ln (\tau)$ values).  We will elucidate the reasons for such a choice in the next section.  Essentially each of the two superimposed Fermi-like functions represents a corresponding regime of light scattering mechanisms.  Figures 2-4(b) show the plot of $1/T_F $ (which is equivalent to $e^{-\epsilon_F}$) versus the blood flow velocity with different tube diameters.  Amazingly, we obtain a simple linear relation between $1/T_F$ and the blood speed.  More interestingly, we have found that the $1/T_F$ versus speed curve is independent of the tube diameter.  As shown in Fig. 5, the three curves of $1/T_F$versus speed measured from different tube diameters completely overlap and can be represented by one simple relation independent of the tube diameter. This result suggests that our measurement setup can potentially measure the blood speed in different artery sizes without having to know the exact dimension of the blood vessel diameter.  This discovery is very important in practical applications because it shows that from the $g_1(\tau)$ curve, which can be obtained from the correlation of the photocurrent, we can obtain the speed of the blood directly for different arteries at a given position without any prior knowledge of the anatomy of the blood vessel.  The discovery of this important relation requires a sound physical foundation, to rule out the possibility for being simply coincidental.  The physical model and mathematical analysis will be discussed next.
\begin{figure}[h!]
\centering\includegraphics[width=13cm]{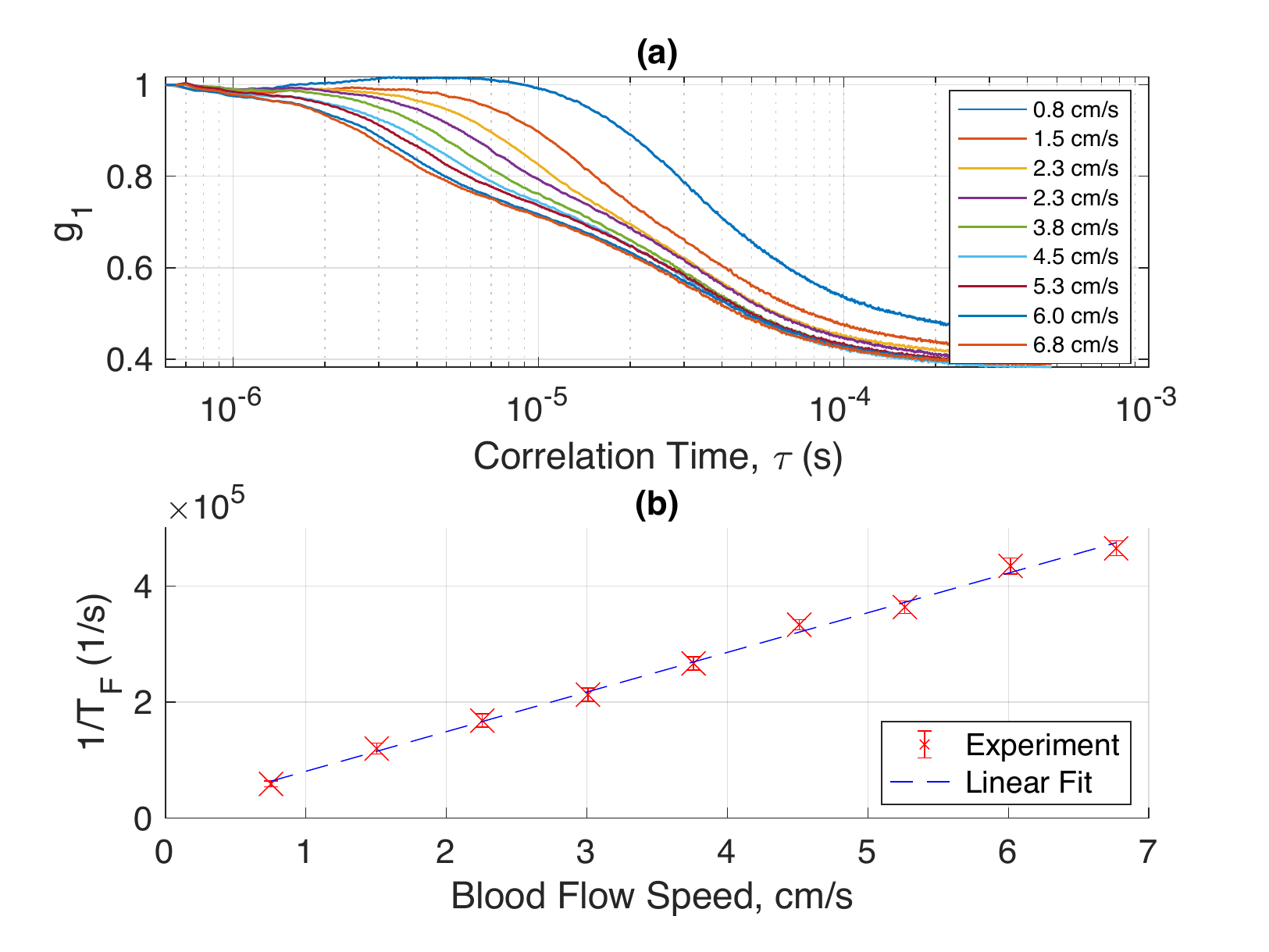}
\caption{(a) Measured E-field correlation function ($g_1$) versus flow speed of whole blood for a tube diameter of 1.68mm.  (b) By taking the inverse of the correlation time at the first inflection point in (a), the decorrelation rate ($1/T_F$) is plotted versus the flow speed.  The scattered data set with error bars represents the 95$\%$ confidence interval over 25 measurements, each for a duration of 1 second. The dashed line is a linear fit of the measured data.}
\end{figure}

\begin{figure}[h!]
\centering\includegraphics[width=13cm]{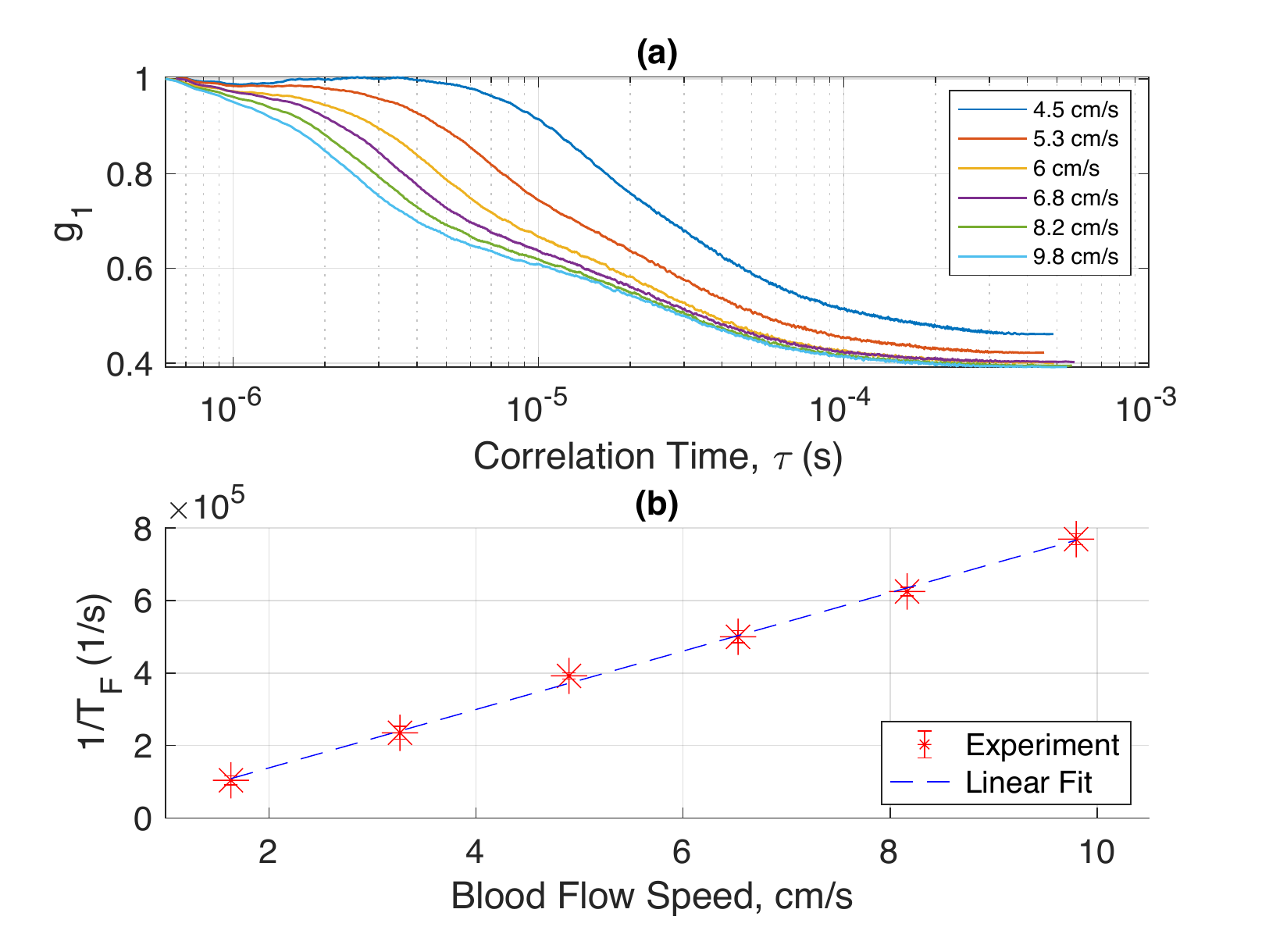}
\caption{(a) Measured E-field correlation function ($g_1$) versus flow speed of whole blood for a tube diameter of 1.14mm.  (b) By taking the inverse of the correlation time at the first inflection point in (a), the decorrelation rate ($1/T_F$) is plotted versus the flow speed.  The scattered data set with error bars represents the 95$\%$ confidence interval over 25 measurements, each for a duration of 1 second. The dashed line is a linear fit of the measured data. }
\end{figure}

\begin{figure}[h!]
\centering\includegraphics[width=13cm]{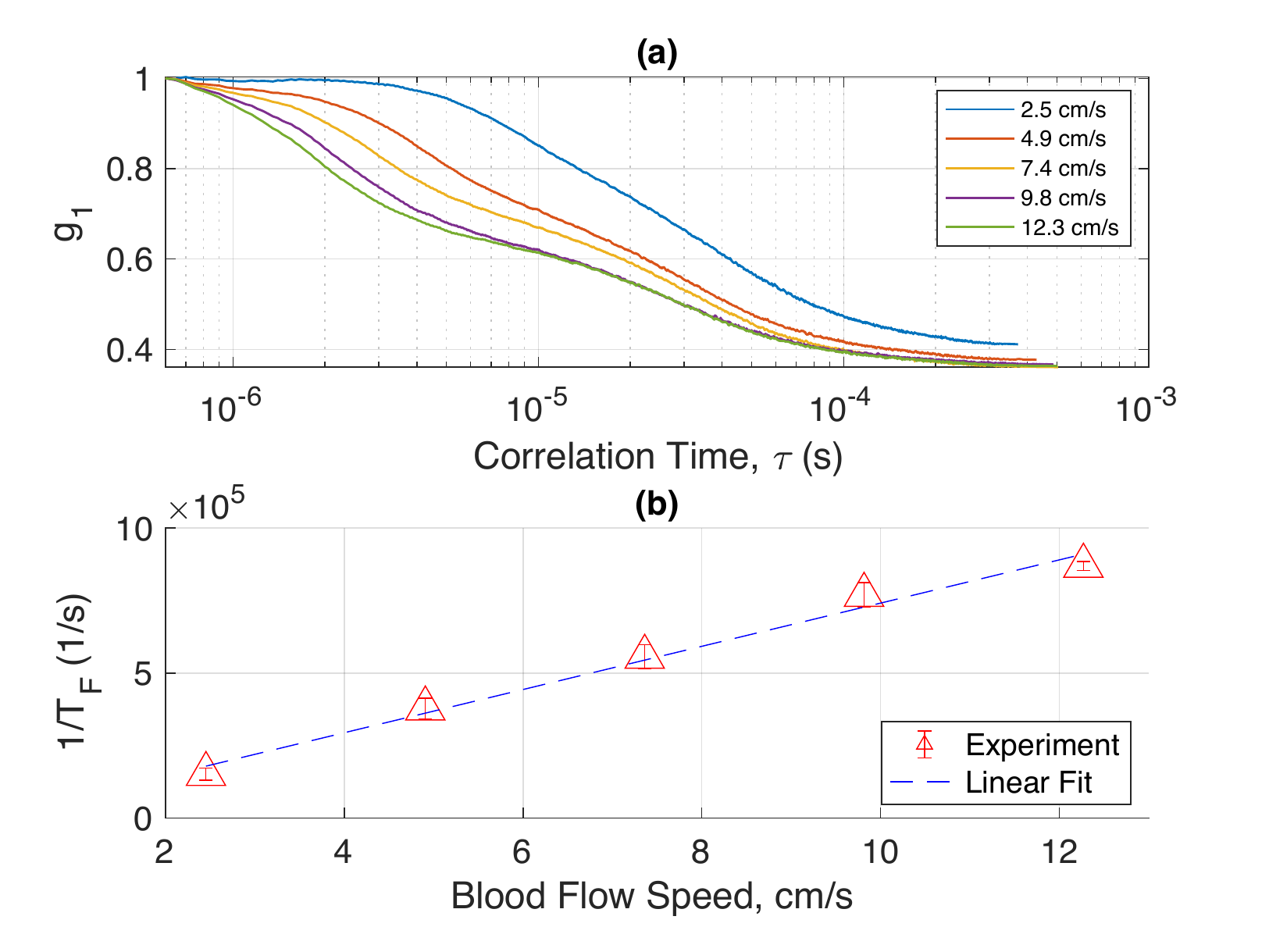}
\caption{(a) Measured E-field correlation function ($g_1$) versus flow speed of whole blood for a tube diameter of 0.93mm.  (b) By taking the inverse of the correlation time at the first inflection point in (a), the decorrelation rate ($1/T_F$) is plotted versus the flow speed.  The scattered data set with error bars represents the 95$\%$ confidence interval over 25 measurements, each for a  duration of 1 second. The dashed line is a linear fit of the measured data. }
\end{figure}

\begin{figure}[h!]
\centering\includegraphics[width=13cm]{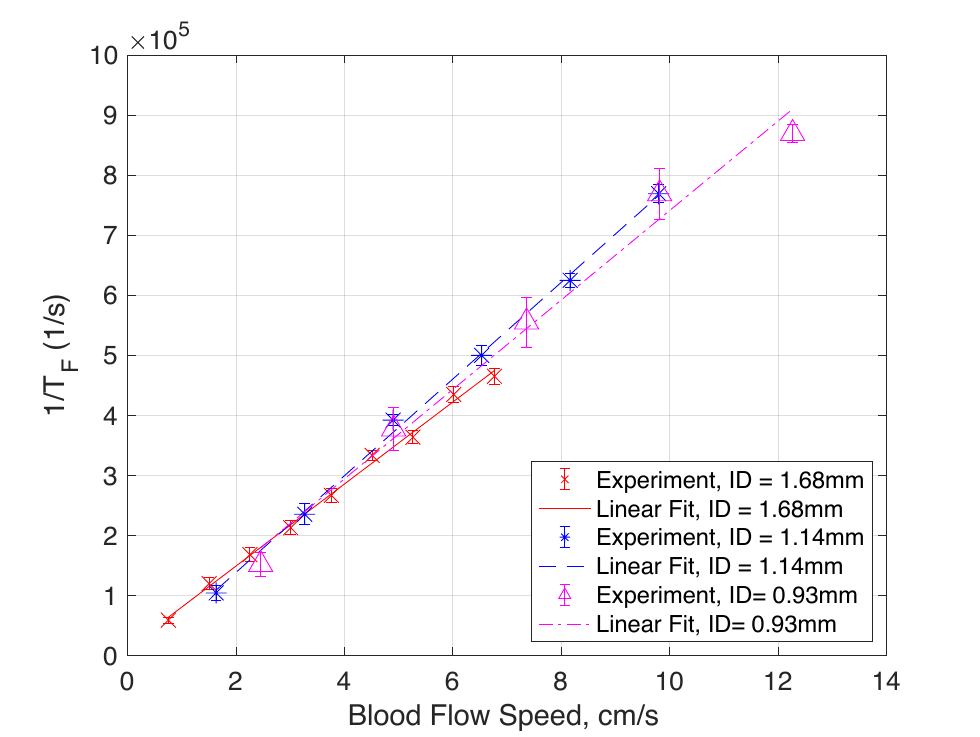}
\caption{Combined decorrelation rate ($1/T_F$) versus blood flow speed plot for different tube ID. The scattered data set with the error bar represents the 95$\%$ confidence interval over 25 measurements, each for a duration of 1 second. The line plots are linear fits of the measured data. The fact that plots for different tube ID collapse into one single line demonstrates the existence of a simple relation between the decorrelation rate and the blood speed and the relation is independent of the tube diameter. }
\end{figure}

\section{Physical Models}
\subsection{Diffused Light Equation for Moving Scatters:}
In order to understand the underlining physics of the relation between the inverse of characteristic decorrelation time, $1/T_F$, and the flow speed of the blood, we describe the physical model and mathematical formulation for our experiment in this section.  We will outline the key steps and, through approximations, produce an analytical relation between the blood speed and the decorrelation time.  The detailed numerical results for more general analyses without approximations will be presented in a separate paper.

People usually take two approaches to model light propagation in a strongly scattered biological medium. In one approach, we start with the wave equation and introduce the scattering and absorption characteristics along the optical path. Despite its rigor in the mathematical formulation, to make the result useful, approximations have to be made to make the problem solvable. Twersky's theory and Dyson's equation fall into this category \cite{ishimaru1978wave}. In another approach, we can formulate the problem in a transport equation that deals with photon energy transport.  These two methods eventually give rise to the same result.  In this paper we take the second approach because of its relative simplicity.

The governing equation is a radiative transfer equation (RTE) in the following form:
\begin{equation}
	\frac{dI(\vec{r},\hat{s})}{ds} = -\rho\ \sigma_t I(\vec{r},\hat{s}) + \rho\sigma_s\int_{4\pi}p(\hat{s},\hat{s}')I(\vec{r},\hat{s}')\ d\Omega' +S(\vec{r},\hat{s})
\end{equation}
\begin{figure}[h!]
\centering\includegraphics[width=6cm]{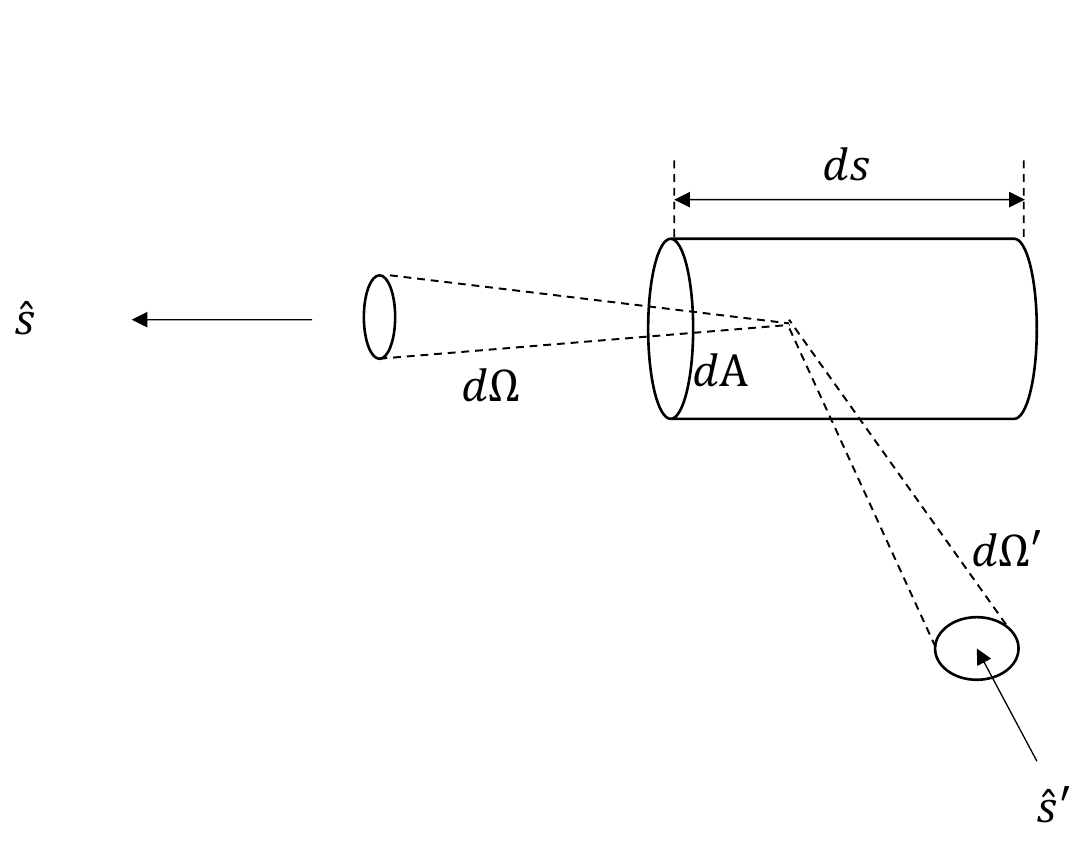}
\caption{Illustration of the elementary scattering volume for Eq. (3). }
\end{figure}

in which $I$ is the light specific intensity with a unit  $Wm^{-2} sr^{-1}$ ,$\rho$ is the particle concentration, $\sigma_s$ is the scattering cross section, $\sigma_t$ is the total scattering cross section which is the summation of scattering cross section and absorption cross section.  $p(\hat{s},\hat{s}')$ is the normalized differential scattering cross section which is sometimes called the phase function and it's unitless, and $S(\vec{r},\hat{s}) $ is the source intensity which has a unit of $Wm^{-3} sr^{-1}$.  Figure 6 illustrates the orientations and coordinates for an infinitesimal scattering volume. Eq. (3) contains 5 coordinates: $x, y ,z,\theta,\phi$ where $x,y,z$ define the position vector $\vec{r}$ of the light intensity, and $\theta,\phi$ represent the beam's propagation direction. The unit vectors $\hat{s}'$ and $\hat{s}$ represent the direction of the incident light and the propagation direction after scattering.

Solving Eq. (3) would produce the full solution for the light scattering problem for any scatter concentration. However, Eq. (3) is very complicated to solve. Fortunately, in most biological samples, the scatter density is so high that photons quickly lose the memory of their path histories after multiple scatterings and it can be justified to assume the light intensity depends on its position $(x,y,z)$ with a slight flux flow in the direction of propagation $(\theta,\phi)$. This would lead to a simplification of the problem, and the average intensity could be described by the diffusion equation which is only dependent on the position $(x,y,z)$.  Mathematically, it means that we can expand the light intensity in spherical harmonics by keeping the zero order and first order term.
\begin{equation}
		I(\vec{r},\hat{s}) \simeq \frac{1}{4\pi}U(\vec{r}) +\frac{3}{4\pi}\vec{F}(\vec{r})\cdot \hat{s}
\end{equation}
\\where the first term is the average intensity and the second term is the small photon flux in the direction of propagation. Applying the diffusion approximation to the RTE, we can obtain a steady state diffusion Eq. (5) to model light propagation in a diffusive medium when the scatters are not moving. As light needs to be scattered multiple times to become diffusive, an important requirement for the diffusion approximation is that the cross section for light scattering is much stronger than the cross section for light absorption. A problem would arise when dealing with light intensity close to the boundary since the light is highly directional at the boundary, violating the diffusive condition. Different boundary conditions have been explored to resolve this problem, including adoption of an extended boundary condition which uses Taylor expansion to convert the Robin boundary condition into a Dirichlet type boundary condition to simplify the solution for Eq. (5) \cite{haskell1994boundary}.
\begin{equation}
    [D\nabla^2-\mu_a]U(\vec{r})= -S(\vec{r})
\end{equation}
\\where $D=1/3\mu_s'$   is the photon diffusivity with a unit of meter and $\mu_s'$ is the reduced scattering coefficient. $\mu_a$ is the absorption coefficient with a unit $m^{-1}$. $S(\vec{r})$ is the source (unit of $Wm^{-3}$) that depends only on the location but not on the propagation direction, different from $S(\vec{r},\hat{s})$ in Eq. (3).When the scatters exhibit motions, the scattered intensity would include time as a parameter.  For statistical optics, it is natural to use field correlation function to capture this dynamic process. Here, we are still talking about the steady state response of the system so that the time dependence can be represented as a parameter in the differential equation.

We start with the case of single scattering event by a moving object. The normalized single scattering function can be written as:
\begin{equation}
    g_1^s(\tau) =\frac{\langle E(t)E^*(t+\tau)\rangle}{\langle E(t)E^*(t)\rangle} = \exp\bigg(-\frac{1}{6}q^2\langle \Delta r_i^2(\tau)\rangle \bigg)
\end{equation}
\\where $g_1^s$ is the single scattering normalized correlation function, $E$ is the scalar electrical field of light, $q$ is the photon momentum transfer by each scattering, $\langle \Delta r_i^2(\tau)\rangle$ is the RMS displacement of the scatters in a duration of $\tau$. The above equation describes the electrical field correlation function due to single scattering. This single scattering function can be incorporated into the radiative transfer equation, yielding the so-called correlation transfer equation which is the dynamic counter part of the static radiative transfer equation \cite{boas1997spatially}:
\begin{equation}
	\frac{dG_1(\vec{r},\hat{s},\tau)}{ds} = -\rho \sigma_t G_1(\vec{r},\hat{s},\tau)\ +\ \rho\sigma_s\int_{4 \pi}p(\hat{s},\hat{s}')\ \ g_1^s(\hat{s},\hat{s}',\tau)G_1(\vec{r},\hat{s},\tau)\ d\Omega' + S(\vec{r},\hat{s})
\end{equation}
\\Similarly, the diffusion approximation can also be applied to obtain the correlation diffusion equation which is again the counterpart of the static diffusion equation described previously. Eq. (8) is the correlation diffusion equation. The diffusion equation can then be solved with appropriate boundary conditions to be compared with experimental results.
\begin{equation}
    [D\nabla^2-\mu_a - \frac{1}{3}\mu_s'k_0^2\langle\Delta r^2(\tau)\rangle]G_1(\vec{r},\tau) = -S(\vec{r})
\end{equation}
\\where $n$ is the refactive index of scattering media, $k_0$ is the wavevector in vacuum, $\mu_s'$ is the reduced scattering coefficient which is the scattering coefficient modified due to scattering anisotropy.  For isotropic scattering, the scattering coefficient would be the same reduced scattering coefficient.
\subsection{Analytic Relation Between Decorrelation Time and RBC Flow Speed}
We assume a plane wave illumination at the surface. Despite the simplification, the approximation can yield a satisfactory result in good agreement with the experiment.  To simplify the practical application, we adopted an experimental design where a single detector covers an area defined by the numerical aperture of 6 hexagonally arranged multimode fiber surrounding the light source (see Fig. 1).  Assuming the blood vessel is cylindrically symmetric, we used a 2D model as shown in Fig. 7.
\begin{figure}[h!]
\centering\includegraphics[width=7cm]{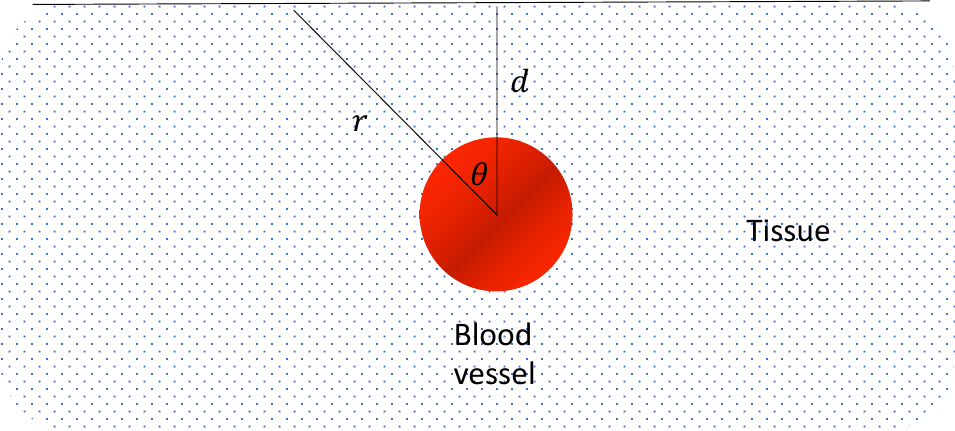}
\caption{Geometry used for theoretical calculation. The circle represents the cross section of blood vessel with parameters: $D_{in}$, $W_1$, $k_1$, $\mu_a^{in}$, $\mu_s^{'in}$; The area outside the circle represents the static scattering media, i.e. tissue with parameters: $D_{out}$, $W_0$, $k_0$, $\mu_a^{out}$, $\mu_s^{'out}$ . We adopt cylindrical coordinates and define the center of the vessel as the origin.}
\end{figure}
We can reduce Eq. (8) into the following diffusion equation,
\begin{equation}
    [\nabla^2 + k^2]G_1(\vec{r},\tau) = -\frac{S(\vec{r})}{D}
\end{equation}
in which
\begin{equation}
    k = jW(\tau)
\end{equation}
\\Since the scattering parameters are different inside and outside the blood vessel, two sets of parameters are used. The parameters inside the blood vessel are represented using subscript 1, i.e. $k_1=jW_1 (\tau)$. The parameters outside the blood vessel are represented using subscript 0, i.e. $k_0=jW_0(\tau)$.
\begin{equation}
    W_1 = \sqrt{\frac{1}{D_{in}}[\mu_a^{in}+\frac{1}{3}\mu_s k_\lambda^2\langle \Delta r^2(\tau)\rangle]}
\end{equation}
\begin{equation}
    W_0 = \sqrt{\frac{\mu_a^{out}}{D_{out}}}
\end{equation}
\\To solve Eq. (9), we need to first find the mean squared displacement $\langle \Delta r^2(\tau)\rangle$. The mean square displacement for RBCs includes Brownian, shear induced diffusion and convective motion. We can represent the position variation of RBCs over a given time interval as\cite{belau2017pulse}:
\begin{equation}
    \langle\Delta r^2(\tau)\rangle = 6D_{\alpha}\tau + \bar{V}^2\tau^2
\end{equation}
where the first term in Eq. (13) is caused by Brownian and shear induced diffusions and the second term by convection.  For RBCs, the diffusion by Brownian motions is much smaller than the shear induced diffusion \cite{wu1990diffusing}. The flow of RBCs inside arteries can be modeled by a laminar flow and the diffusion coefficient can be represented as \cite{owen2007arterial}
\begin{equation}
    D_{\alpha} = \alpha_s\bigg|\frac{\partial v_{RBC}}{\partial r}\bigg| = \frac{4}{3}\alpha_s 
    \frac{V_{max}}{a}
\end{equation}
 $\alpha_s$ is a parameter describing the interaction strength among blood cells due to shear and its value has been measured experimentally \cite{goldsmith1979flow,funck2018characterization,nanne2010shear,higgins2009statistical,biasetti2014synergy,cha2001evaluation,tang2018shear}.The radial dependent shear rate is usually replaced by its average value due to the fact that multiple scattering will lead to an ensemble averaging across all radial locations . For tissue blood perfusion, the slow flow speed and small blood vessel diameter make the diffusive motion the dominant effect compared to the convective motion \cite{sakadvzic2017theoretical}. The situation is reversed, however, for main arteries and our phantom model where the vessel inner diameter is of millimeter size and the blood flow speed is several centimeters per second\cite{gabe1969measurement}.  In such cases, the convective motion becomes the dominant effect. Hence in our analysis, we have ignored the diffusion contribution and included only the convective flow contribution in Eq. (13).

We can then write the general solution of the correlation function in Eq. (9) as follows,
\begin{equation}
    G_1(\vec{r},\tau) = G_1^{in}(\vec{r},\tau) + G_1^{sc}(\vec{r},\tau)
\end{equation}
\\The first term is the inhomogeneous solution and the second term is the homogeneous solution for Eq. (9) in the absence of the source. The diffusion equation under a given boundary condition can be solved in polar coordinates and the method of separation of variables. For an approximate analytic solution, we kept only the zeroth order term since all higher order terms are insignificant compared with the zeroth order term. A continuity boundary condition is applied at the cylinder interface between the blood vessel and the surrounding tissue. The air-tissue surface is ignored for simplicity in the solution. After solving the differential equation with some approximations based on the numerical value of actual tissue and blood cell scattering properties, the normalized correlation function $g_1$ can be written in the form of Eq. (16), inspired from the Fermi-Dirac function in semiconductor physics (i.e. by performing the transformations: $\ln(\tau) = \epsilon, \ln(T_F) = E_F $,Eq. (16) is similar to the Fermi-Dirac function). 
\begin{equation}
    g_1(r,a,\theta,\tau) = \frac{1-g_1(r,a,\theta,\infty)}{1+\frac{\tau}{T_F}}+ g_1(r,a,\theta,\infty)
\end{equation}
\begin{equation}
    \frac{1}{T_F} \approx V n k_0\sqrt{\frac{\mu_s^{'in}}{3\mu_a^{in}}}
\end{equation}
where V is the flow velocity of blood, n is the refractive index of blood, $k_0$  is the wavevector of 784 $nm$ light in vacuum, $ \mu_s^{'in}$ is the reduced scattering coefficient of blood ($m^{-1} $), and $ \mu_a^{in} $ is the absorption coefficient of blood ($m^{-1}$). If we define $T_F$ as the characteristic decorrelation time, and $1/T_F$ as the characteristic decorrelation rate, we obtain the key relation that the characteristic decorrelation rate is only a function of flow speed and blood optical parameters, as shown in Eq. (17).  In Eq. (16), $g_1(r,a,\theta,\infty)$ is the asymptotic value of $g_1$ when $\tau$ approaches infinity, and its value is a function of detector position and blood vessel diameter. In a separate study that solves Eq. (9) numerically, we have validated the approximations that led to the analytic solution for $g_1$ in Eq. (16) and most importantly, the linear relation between $1/T_F$ and the blood flow velocity. We have also shown that the decorrelation time for 2D and 3D analyses is nearly the same. The detailed mathematical derivations as well as the numerical computations for the diffused light model will be described in a separate publication.

From the approximated relation in Eq. (17), we find that the characteristic decorrelation rate is proportional to the flow speed.  The proportional constant has a square root dependence on the ratio between the scattering coefficient and the absorption coefficient. This is intuitive since scattering events cause dephasing, and light absorption would terminate the scattering process.  
\section{Discussions}
The measured characteristic decorrelation rate ($1/T_F$) is obtained by calculating the first inflection point of the curves in Figure 2-4. The behavior of the curves at longer time is more complicated, and can be explained by other slower decorrelation processes than convection such as shear induced diffusion, thus yielding two superimposed Fermi-like functions as shown in Figure 2-4.  To use the model described in the previous section, we need to focus on the short time behavior driven by convection. Figure 8 shows the calculated and measured characteristic decorrelation rate under different flow speeds and tube diameters. An excellent agreement between theory and experiment was achieved, confirming that in the regime where convective flow is the dominating factor for decorrelation, the characteristic decorrelation rate is proportional to the blood speed and independent of the vessel diameter.
\begin{figure}[h!]
\centering\includegraphics[width=12cm]{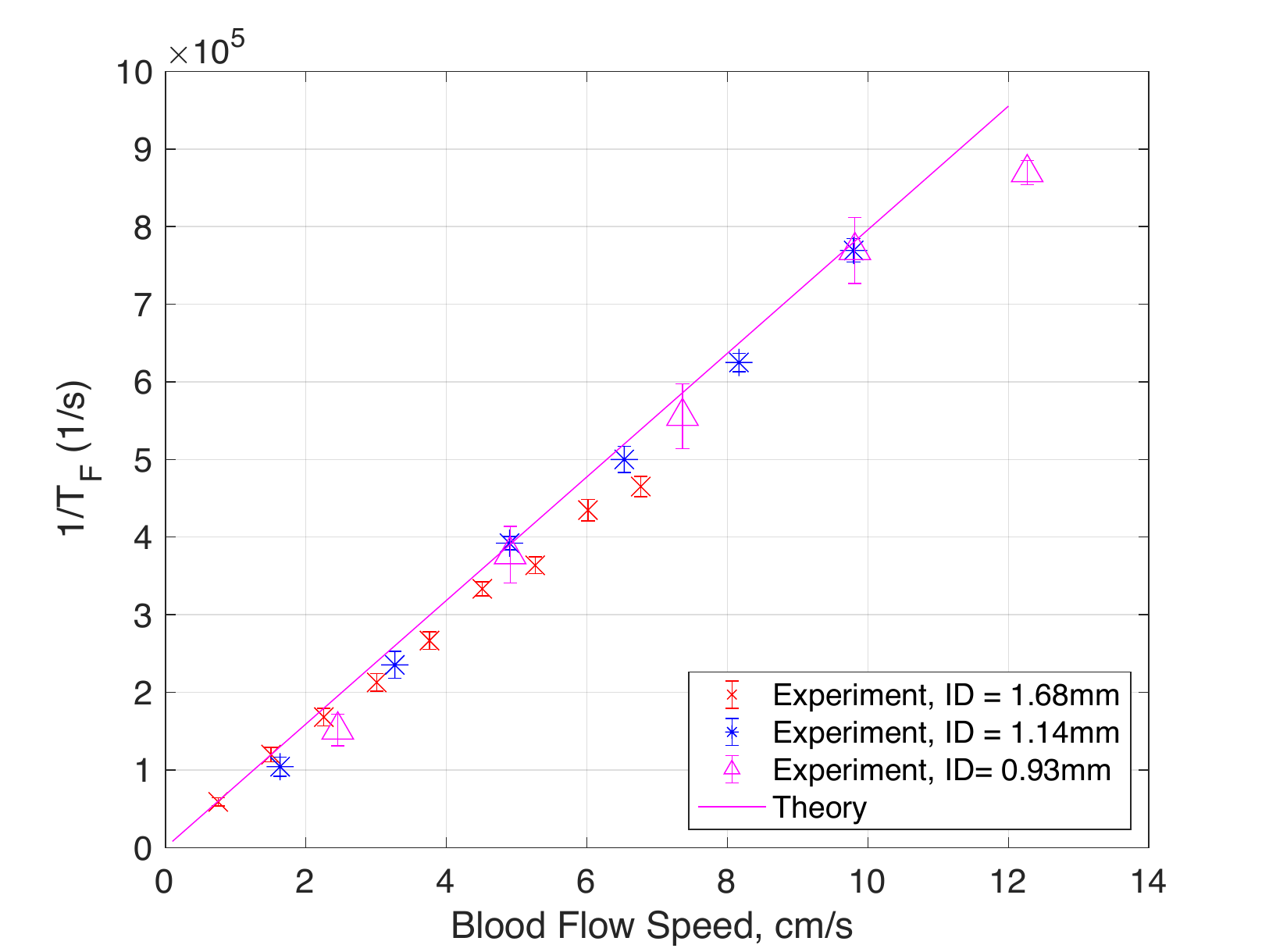}
\caption{Comparison between theoretical calculations (from Eq. 17) and experimentally measured decorrelation rates under different blood flow speeds and tube diameters. The scattered data set with the error bar represents the 95$\%$ confidence interval over 25 measurements.  Each measurement took 1 second. The following parameters are used in the calculations: $n=1.36$, $\mu_s^{'in}=1600 m^{-1}$and $\mu_a^{in}=1000 m^{-1}$ for deoxygenated blood\cite{lazareva2018blood,li2000refractive,bosschaart2014literature,jacques2013optical}.}
\end{figure}
\section{Conclusion}
In this work, we demonstrated a simple technique to directly measure the blood flow speed in main arteries based on the diffused light model.  The device uses a single fiber bundle, a diode laser, and a photoreceiver.  The concept is demonstrated with a phantom that uses intralipid hydrogel to model the biological tissue and an embedded glass tube with flowing human blood to model the blood vessel.  The correlation function of the measured photocurrent was used to find the electrical field correlation function via the Siegert Relation.  Interestingly the measured electric field correlation function $g_1(\tau)$ and $\ln (T_F)$ shows a relation similar to the Fermi-Dirac function, allowing us to define the $ln(\tau)$, equivalent to the ``Fermi energy'' occurring at the first inflection point of $g_1(\tau)$.  Surprisingly, the value $1/T_F$, which we call characteristic decorrelation rate, is found to be linearly proportional to the blood speed and is independent of the diameter of the tube diameter over the size and speed ranges for major arteries.  This striking property can be explained by an approximate analytic solution for the diffused light equation in the regime where the convective flow would dominate the decorrelation.  This discovery is highly significant because, for the first time, we can use a simple device to directly measure the blood speed in major arteries without any prior knowledge or assumption about the geometry or mechanical properties of the blood vessels.  A non-invasive method of measuring arterial blood speed produces important information about health conditions.  Although the device and setup can be further optimized (e.g. adding different light wavelengths) and the physical model can be expanded to acquire more information from the measurements, the work has paved its way to a new promising modality for measurements of blood supplies to vital organs.




\bibliography{references}






\end{document}